\documentclass[%
 reprint,
 amsmath,amssymb,
 aps,
prb,
twocolumn,
]{revtex4}

\usepackage{graphicx}
\usepackage{dcolumn}
\usepackage{bm}

\begin{document}

\preprint{APS/123-QED}

\title[]{Dynamic cloaking of a diamond-shaped hole in elastic plate}
\author{Kun Tang$^{1, 2}$, Eitam Luz$^1$, David Amram$^1$, Luna Kadysz$^1$, S\'ebastien Guenneau$^3$ \& Patrick Sebbah$^1$}

\affiliation{$^1$ Department of Physics, The Jack and Pearl Resnick Institute for Advanced Technology,
Bar-Ilan University, Ramat-Gan 5290002, Israel}
\affiliation{$^2$ Research Center for Humanoid Sensing, Zhejiang Lab, Hangzhou 311100, China.}
\affiliation{$^3$ UMI 2004 Abraham de Moivre-CNRS, Imperial College London, London SW7 2AZ, United Kingdom.}

\begin{abstract}
{Invisibility} cloaks for flexural waves have been mostly examined in the continuous-wave regime, while invisibility is likely to deteriorate for short pulses. Here, we propose the practical realization of a unidirectional invisibility cloak for flexural waves based on an area-preserving coordinate transformation. Time-resolved experiments reveal how the invisibility cloak deviates a pulsed plane-wave from its initial trajectory, and how the initial wavefront perfectly recombines behind the cloak, leaving the diamond-shaped hole invisible, notwithstanding the appearance of a forerunner.
Three dimensional full-elasticity simulations support our experimental observations.
\end{abstract}

\maketitle


In 2006, Leonhardt \cite{Leonhardt1777} and Pendry et al. \cite{Pendry1780} proposed independently the idea of invisibility cloak for electromagnetic radiation. The former proposal is based on some conformal mapping while the latter makes use of the form-invariance of Maxwell's equations under geometric transforms and consists in stretching a small hole to create a cloaked volume surrounded by an anisotropic heterogeneous shell that is impedance-matched to free space. The latter route is more general and should make the hole and what it might contain invisible to electromagnetic radiations beyond the limit of geometric optics.
This concept has been validated experimentally for monochromatic electromagnetic waves \cite{Schurig977}. However, because the cloak's performance highly depends upon the dispersion properties of its metamaterial's constituents that approximate the required optical parameters, it seems hopeless to achieve passive broadband optical cloaking due to the light speed limit \cite{Milton_quasistatic}. This fundamental limit can be overcome by considering other types of waves with lower velocity such as
acoustic waves \cite{Cummer_2007}, hydrodynamics \cite{hydrodynamic_cloak} and water waves \cite{water_wave_cloak}. Such waves are governed by form-invariant equations. This is not true anymore for the elasticity equations that transform into the so-called Willis's equations \cite{Milton_2006,Norris_2011,Achaoui2020}, except in the frameworks of Cosserat elasticity \cite{Nassar_PRSA,Nassar_PRL,brun_apl}, where the transformed elasticity tensor does not have the minor symmetry. Consequently, if cloaking is possible for such class of elastic waves, it would be of a different nature than its acoustic and electromagnetic counterparts. To simplify the problem, researchers resort to studying the particular case of flexural waves in thin plates, which are described by the more tractable fourth-order Kirchhoff-Love equation.

There have been various theoretical proposals for the design of elastic invisibility cloaks for flexural waves \cite{PRB_farhat,PRL_farhat,COLQUITT2014131,Brun_2014,PRE_darabi,arxiv_pomot}, followed by their experimental validations \cite{cloaking_wegner,cloaking_darabi,Misseroni2016}.
However, experimental investigations of cloaking have been mostly restricted to continuous-wave (CW) excitation \cite{Schurig977,optical_cloaking,cloaking_wegner,cloaking_darabi,2D_ground_cloaking,Cloaking_nico_fang}. Response of invisibility cloaks to short pulses has been barely tested \cite{pulse_LC,2D_ground_cloaking,PNAS_surface_wave,SR_surface}, since inhomogeneous, magnetic or anisotropic metamaterial structures are subject to inherent spectral dispersion, which is likely to distort the pulse both in space and time, making its reconstruction challenging \cite{optic_delay,temporal_boris}. Cloaking over a broad frequency range has been realized in various systems, including acoustic \cite{Cloaking_nico_fang}, elastic \cite{cloaking_wegner,cloaking_darabi}, and water waves \cite{water_wave_cloak}. But achieving broad spectral range operation does not guarantee that a pulse propagating through the transformed medium will remain undistorted.

\begin{figure}
\centering
\includegraphics[width = 8cm]{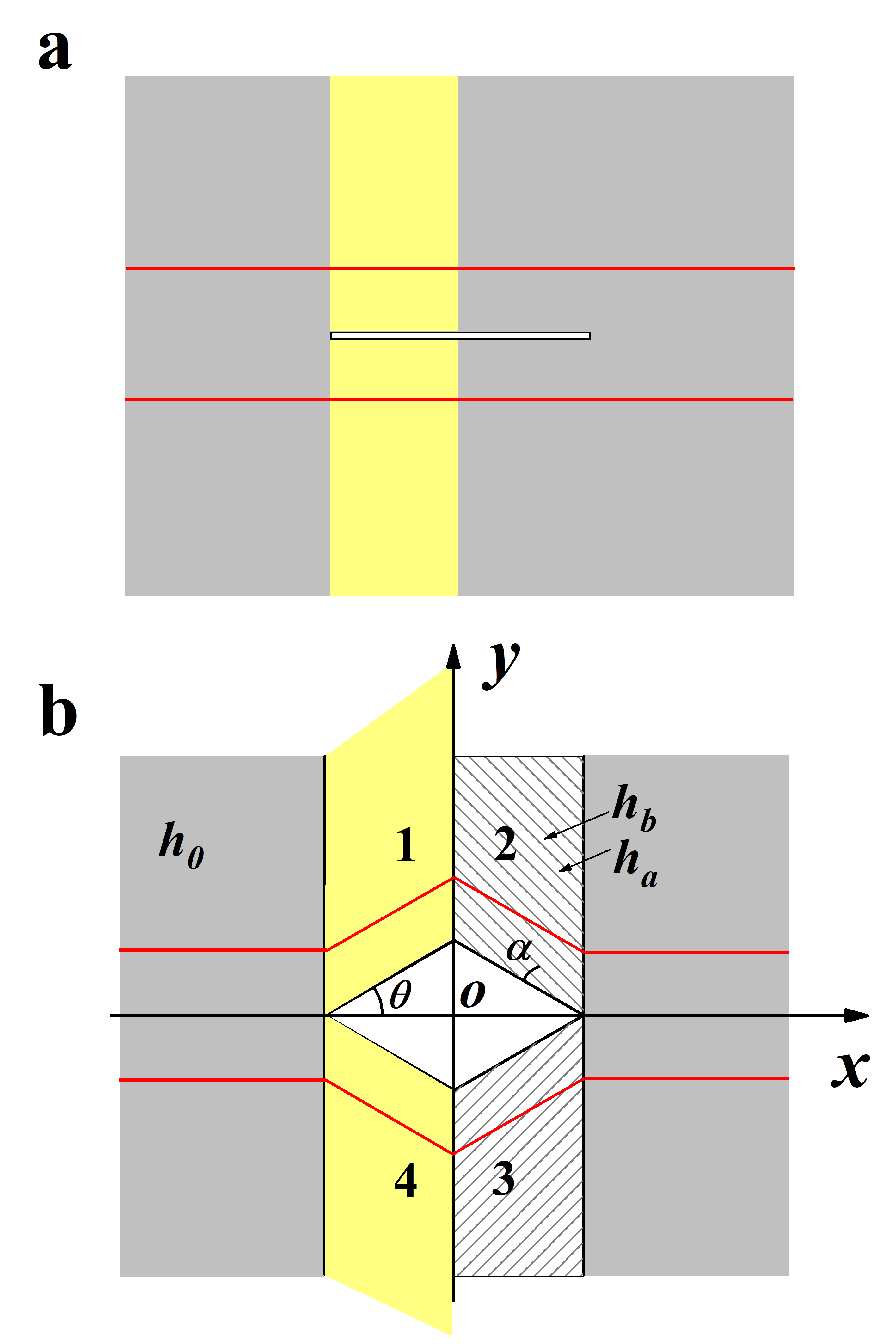}
\caption{(a) Schematics of a free plate with a stress-free segment. (b) Schematics of the proposed unidirectional cloak for a diamond-shaped defect. The cloak is made of four blocks (labeled with 1 to 4) of alternating layers with thicknesses $h_a$ and $h_b$, which form an angle $\alpha$ with the edge of the diamond-shaped defect. The red lines illustrate the longer trajectories of the flexural waves propagating through the cloaked region.
The mapping of the yellow slab in (a) onto the yellow cloaking region in (b) preserves area. The cloak extends infinitely along the $y$-direction.
}
\end{figure}

The coordinate transformation used in the design of electromagnetic cloaks with cylindrical geometry leads to a gradient distribution of permittivity and permeability, which necessitates engineering the magnetic response of materials hardly available in the optical range \cite{Pendry1780,Schurig977}. To solve this issue, a reduced set of parameters has been proposed, which provides with non-magnetic structures, while preserving the cloaking performances, but which suffers from reflection and scattering due to impedance mismatch at the cloak's outer boundary \cite{optical_cloaking}. Another approach to avoid magnetic materials and preserve the invisibility of the device itself is to maintain the determinant of the Jacobian transformation tensor at unity, therefore preserving the volumes throughout the space \cite{Non_magnetism,han_ol,xu_josa,han_oe}. Compared to cylindrical cloaking, this so-called non-magnetic geometrical transformation is continuous i.e. adiabatic \cite{shifter_huanyang}. The space is not abruptly stretched or compressed and the topology is conserved during the transformation process \cite{shifter_rahm}, ensuring perfect impedance matching at the boundaries. This volume-preserving method has recently been considered in the design of waveshifter and rotator for elastic waves \cite{kun_AFM}, where it could meet the challenge of designing intrinsically reflection-less elastic devices. \\
In this Letter, we adapt the non-magnetic geometrical transformation to elastic waves and call on for its continuous and area preserving character to design a unidirectional cloak for flexural waves. Here, we investigate the pulse dynamics of flexural waves propagating inside a 3D-printed cloaked plate by mapping the spatio-temporal field distribution in response to a short pulse excitation. The elastic pulse is shown to break apart and perfectly follow the $\pm 30^{\circ}$ bent edges of the cloak with minor back reflections, and converge back to a plane wavefront, leaving the diamond-shaped hole undetected. To the best of our knowledge, diamond-shaped cloaking of elastic waves has never been experimentally demonstrated, especially under the incidence of short pulses \cite{cloaking_wegner,cloaking_darabi}. Three dimensional full-elasticity simulations support our experimental observations.

We consider a plain thin plate that supports flexural waves. Suppose a narrow slit is perforated in the plate, as in Fig.~1a. This slit will leave unaffected a plane-wave incident along the slit direction. Suppose now that we open the slit into a diamond-shaped hole by simply pulling each edge of the slit at the middle. Such an operation will inevitably deform the surroundings. Instead of stretching or compressing the space around the hole, we choose rather to shift it above and below the hole along the vertical axis, in order to preserve the total area, as shown in Fig.~1(b). Theoretically, each deformed section should extend to infinity. Here, we will use PML boundary conditions in numerical simulations and absorbing material in experiments to mimic an infinitely extended medium.
To operate this area-preserving transformation, we consider a point-to-point mapping of the coordinates from a horizontal region in virtual space $(X,Y)$ (yellow rectangle in Fig. 1(a)) into an oblique region in the real space $(x,y)$ (yellow rhomboid in Fig. 1(b)):
\begin{equation}
\centering
     \begin{cases}
      x  = X  \\
      y  = X \tan\theta + Y  \\
       \end{cases}
       \label{eq:waveshifter}
\end{equation}
where $\theta$ is the angle, in the direction of which the wavefront is bent. The Jacobian matrix of the above geometrical transformation is
\begin{equation}
  F =\left(
 {\begin{array}{ccccc}
   1 & 0\\
   t & 1\\
  \end{array}}
       \right)
\label{eq:waveshifter_jaco}
\end{equation}
where $t= \tan\theta$. Note that $J=\mid {\rm det}\, F \mid = 1 $, which refers to an area preserving transformation. In contrast to previous transformations for flexural wave cloaking \cite{PRB_farhat,PRE_darabi}, the off-diagonal term $t$ is not zero.\\
This geometric transformation is readily realized in a waveshifter, where the wave follows a bent waveguide, while its wavefront remains unchanged \cite{kun_AFM}.
Here we design a thin plate with a cloak inspired by the waveshifter \cite{shifter_huanyang}. Four bending regions ($\pm \theta$) are arranged in a symmetric fashion, leaving a bare diamond-shaped hole as the cloaking region (Fig. 1(b)).
The wavefront of an incident plane wave splits in two, each half taking mirror symmetric directions, before recombining on the other side of the cloaked region, as shown by the red lines in Fig. 1(b).

In the tracks of Ref \cite{PRL_farhat,COLQUITT2014131,Brun_2014,PRE_darabi,arxiv_pomot}, we assume that the transformed Kirchhoff-Love equation can be interpreted in terms of an anisotropic rigidity of the following form
\begin{equation}
  \overline{D}
       = D_0 F F^T F F^T J^{-2}
\label{eq:waveshifter_trans}
\end{equation}
Since $J=1$, the plate density is not affected by the transform. We further note that in general there is one-to-one correspondence between an anisotropic heterogeneous Kirchhoff-Love plate equation, and an anisotropic heterogeneous von-Karman equation (the latter has pre-stress terms), but here the Hessian of the transformation vanishes and thus the Kirchhoff-Love plate equation is effectively form invariant \cite{arxiv_pomot,golgoon2021transformation}.

A medium with this particular anisotropic rigidity $\overline{D}$ can then be realized with a subwavelength structure, by invoking effective medium theory. Here we propose to approximate the homogeneous anisotropic medium by a bi-layered structure, composed of alternating layers of identical width but different rigidity $D_A$ and $D_B$. The effective rigidity tensor for an obliquely layered system with a rotation angle $\alpha$ can be represented by
\begin{equation}
  \overline{D}_{oblique}
  = R_{\alpha}^T \left[
 {\begin{array}{ccccc}
   \ D_{\parallel} & 0\\
   0 & D_{\perp} \\
  \end{array}}
       \right] {R_{\alpha}}
       \label{eq:waveshifter_obliq}
\end{equation}
with $D_{\parallel}=2D_A D_B/(D_A+D_B)$, $D_{\perp}=(D_A+D_B)/2$ and $ R_{\alpha} = \left[
 {\begin{array}{ccccc}
   \cos\alpha & -\sin\alpha\\
   \sin\alpha & \cos\alpha\\
  \end{array}}
       \right] $, the conventional rotation matrix.\\

Identifying Eq.~(3) and Eq.~(4), we obtain the rigidity profile for each layer, which can be easily realized by alternating the plate thickness in regions $A$ and $B$, between $h_A$ and $h_B$ around a background thickness $h_0$. Since the density of the cloak is equal to that of the surrounding plate, impedance matching is naturally satisfied at the interface between layered medium and surrounding plate.

\begin{figure}
\includegraphics[width = 8.5cm]{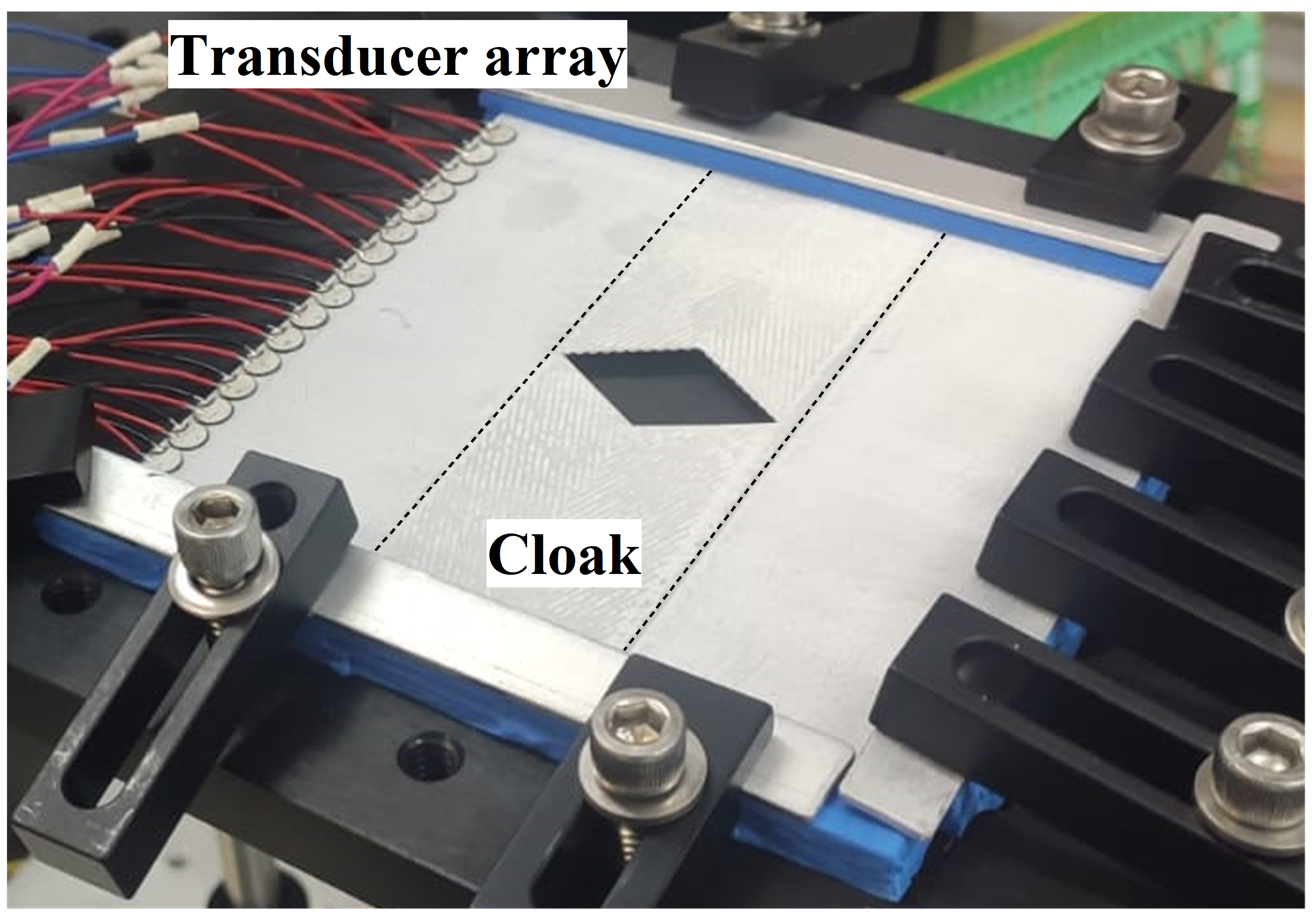}
\caption{Experimental system. The 3D printed ceramic plate is a $h_0=$0.3~mm thick, 100~mm$\times$100~mm square. A diamond-shaped hole is perforated in the center. The cloak region is delimited by the dashed lines. Sixteen piezoelectric diaphragms are bonded along one edge of the plate to generate a one-directional Gaussian pulsed plane wave.}
\label{setup}
\end{figure}

\begin{figure}
\includegraphics[width = 7.2cm]{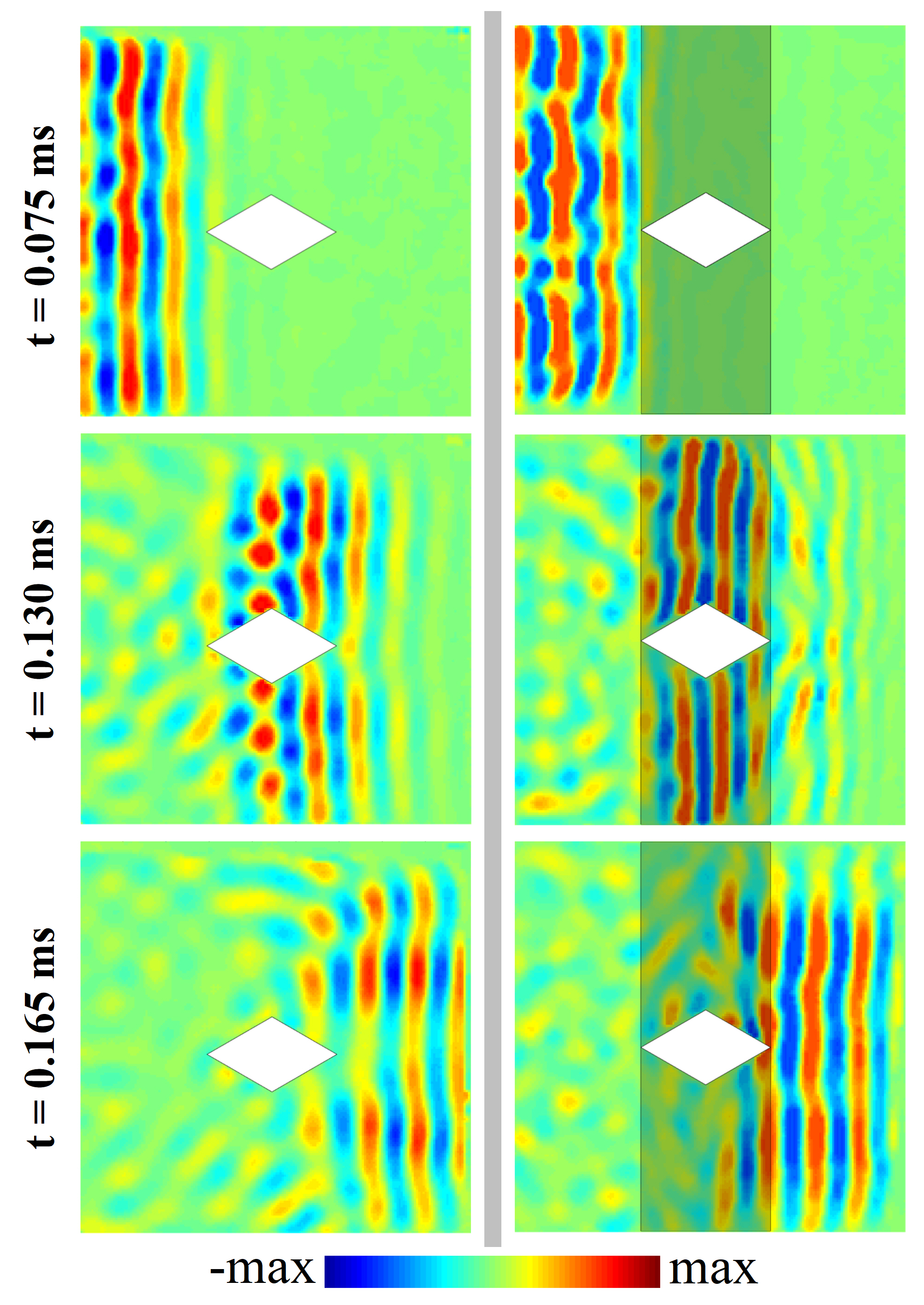}
\caption{Experiment: Velocity-field distribution measured at three different times $t$ = 0.075 ms (upper panel), 0.130 ms (middle panel), and 0.165 ms (lower panel), for (left) the plain plate with a diamond-shaped hole and (right) the plate with a cloak surrounding the hole. The initial 12.5~$\mu$s gaussian pulse at $t$ = 0 is a plane wavefront, at central frequency, 40 kHz.}
\label{experiment}
\end{figure}

The aforementioned four bending regions are arranged in a symmetric way and form a metamaterial cloak surrounding a diamond-shaped hole. Its implementation is sketched in Fig. 1(b) (left side of the hole).

This cloaking geometry, which was proposed earlier for monochromatic light waves \cite{shifter_huanyang}, is tested here for pulsed elastic waves.
We use the parameters calculated earlier \cite{kun_AFM}:
$\theta=30^\circ$, $h_a=0.17~mm$, $h_b=0.54~mm$, $h_0=0.3~mm$, and $\alpha=37^\circ$.
Two 0.3~mm-thick, 100~mm$\times$100~mm-long ceramic plates with a diamond-shape hole have been 3D-printed, one with and one without the cloaking structure.
To design the system, we used 3D-printing based on nanoparticle jetting technology (Xjet, Carmel 1400) and Zirconia (ZrO2), a ceramic with the following elastic parameters: Young’s modulus $E$=207~GPa, mass density $\rho$=6040~kg.m$^{-3}$, and Poisson’s ratio $\nu$=0.32.
Each plate is sandwiched at its four edges between two 3~mm-thick layers of blu-tack. To maximize absorption at the edges, external clamps have been used to maintain the absorber in tight contact with the plate (see Fig.~\ref{setup}). Sixteen piezoelectric ceramic disks (Steminc SMD05T04R111WL, 5 mm-diam., resonant frequency 450kHz) have been glued along one edge of the plate and are driven individually by a 32-channel high-density analog output (AO) module (NI PXIe-6738) coupled to a 32-channel amplifier (PiezoDrive PD32). The amplifier balances each channel to compensate for efficiency variations observed between transducers and yield a plane wave. The AO module generates a 12.5~$\mu$s Gaussian pulse, with carrier frequency 40 kHz. The laser head of a laser vibrometer (Polytec sensor head OFV-534, controller OFV5000) is x-y scanned in 1~mm steps at a distance 30~cm from the flat surface of the plate. The images are processed using a median filter and a cubic interpolation.

The spatio-temporal evolution of the velocity field is recorded for the cloaked plate and the reference plate and shown in two movies available in the \emph{Supplementary Information}. Three snapshots are selected and shown in Fig.~\ref{experiment}(right column) at early time when the pulsed wavefront forms ($t$ = 0.075 ms), when it crosses the free-standing obstacle ($t$= 0.130 ms) and at later time, when it emerges on the other side ($t$ =0.165 ms). In the cloaked system, the pulse wavefront is first seen to split, then to follow up and down the edges of the diamond hole, before it recombines on the other side and reforms into a unique plane wavefront. This demonstrates that the hole is invisible to the elastic pulse. This is in contrast with the forward diffraction pattern observed with the reference plate, in absence of cloak Fig.~\ref{experiment}(left column). Interestingly, the cloak is impedance-matched, as no reflection occurs at its interface. The residual speckle noise is attributed to imperfect excitation (contributions away from the incident direction). Also, the cloak works remarkably well on both sides of the diamond where the wavefront direction is nicely maintained, while it is fragmented in absence of cloak. This  realizes the desired mapping from an horizontal into a bent direction, and reciprocally. We further note the presence of a precursor wavefront (akin to a Brillouin-Sommerfeld forerunner \cite{sommerfeld1914fortpflanzung,brillouin1914fortpflanzung}) after the cloak at $t=0.130$ ms, which is generated by the dispersion induced by the layered medium, as predicted theoretically for a similar transformed medium in the electromagnetic wave framework in \cite{temporal_boris}.

To model the experiment, we use finite element method using COMSOL Multiphysics. The simulations of the pulse response for the metamaterial plate hereafter are done with the full-3D layered structures, rather than with the simplified Kirchoff-Love plate equations.
The incident plane wave is a Gaussian pulse with full width at half maximum(FWHM) around 19.5$\mu s$, centered at frequency 40kHz.

\begin{figure}[h!]
\includegraphics[width = 8cm]{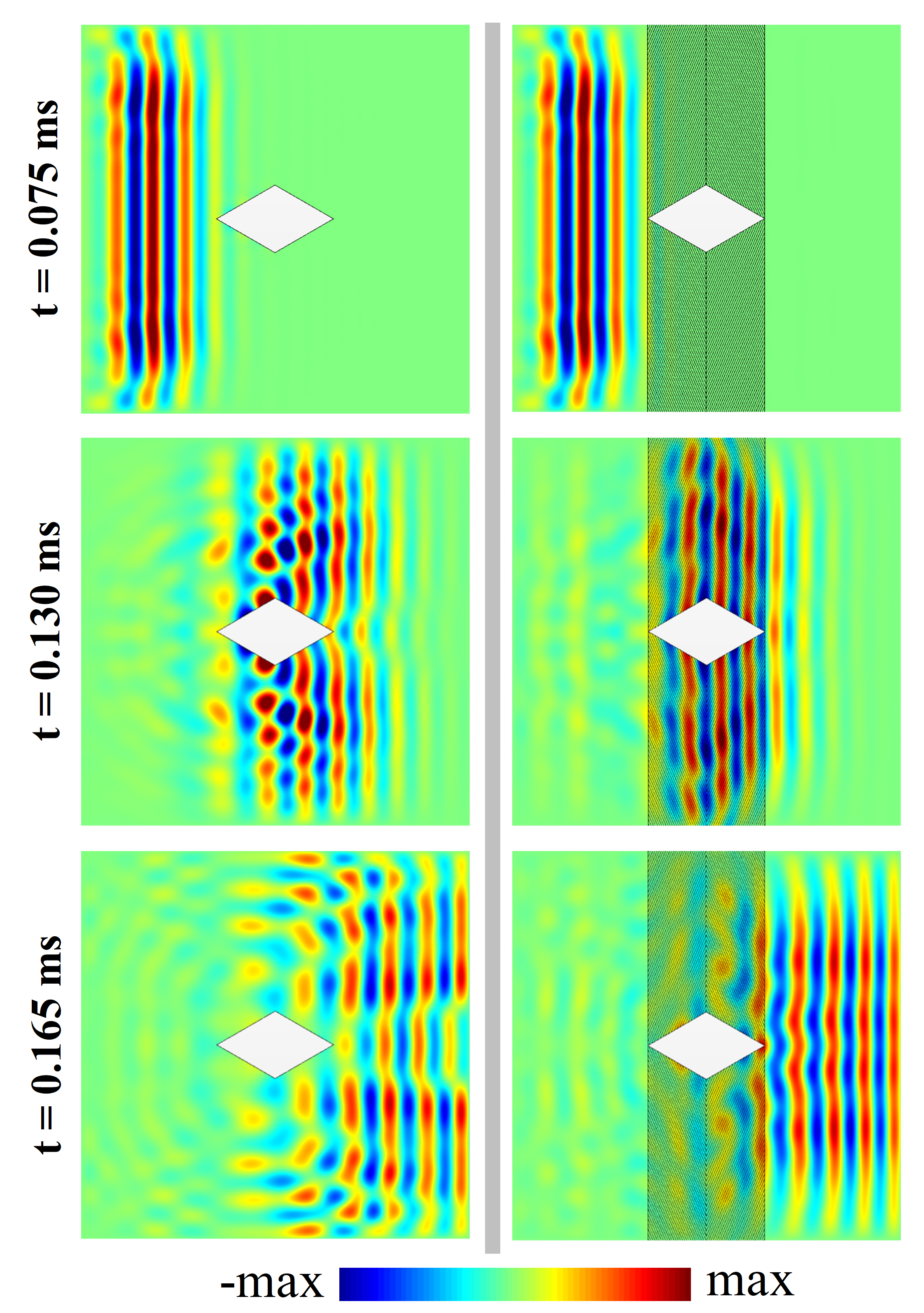}
\caption{Numerical simulations: Velocity-field distribution computed at three different times $t$ = 0.075 ms (upper panel), 0.130 ms (middle panel), and 0.165 ms (lower panel), for (left) the plain plate with a diamond-shaped hole and (right) the plate with the proposed cloak surrounding the hole. The initial plane wavefront is a Gaussian pulse line.}
\label{simulations}
\end{figure}

Figure \ref{simulations} compares the field distribution of the vertical elastic velocity resulting from the scattering by a bare hole with stress-free boundaries without and with the cloaking device. The velocity field is shown at three different time steps. In absence of cloak, the plane wavefront splits as it hits the hole and then rapidly breaks as a result of multiple interferences and diffraction (Fig.~\ref{simulations}(left column)). With the cloak, however, the flat wavefront is maintained while following the edges of the obstacle and recombines nicely after the hole (Fig.~\ref{simulations}(left column)). Beyond the hole, a perfectly flat wavefront is restored at $t=0.165$ ms, consistently with experiments at the same time step in Fig. \ref{experiment}. Remarkably, in addition to the invisibility effect, the cloak itself is invisible, with negligible back reflection. We note nonetheless the presence of a forerunner beyond the cloak at $t=0.130$ ms, consistently with experiments at the same time step in Fig. \ref{experiment}. We checked that the temporal spreading of the pulse observed after the obstacle is solely due to the natural dispersion of the flexural waves, as it would occur naturally in a plain plate without the cloak.
Even if the pulse is not distorted, there should be however a measurable delay of the pulse in forward scattering for the cloaked obstacle, in comparison with a bare plate (without hole and cloak device).
This is confirmed by examining the spectral domain at a single frequency 40kHz. A phase delay after restoration of the wavefront is observed when comparing the cloaked plate and the bare plate (see Figure S1 in Supporting Information).

\begin{figure}[h!]
\includegraphics[width = 8cm]{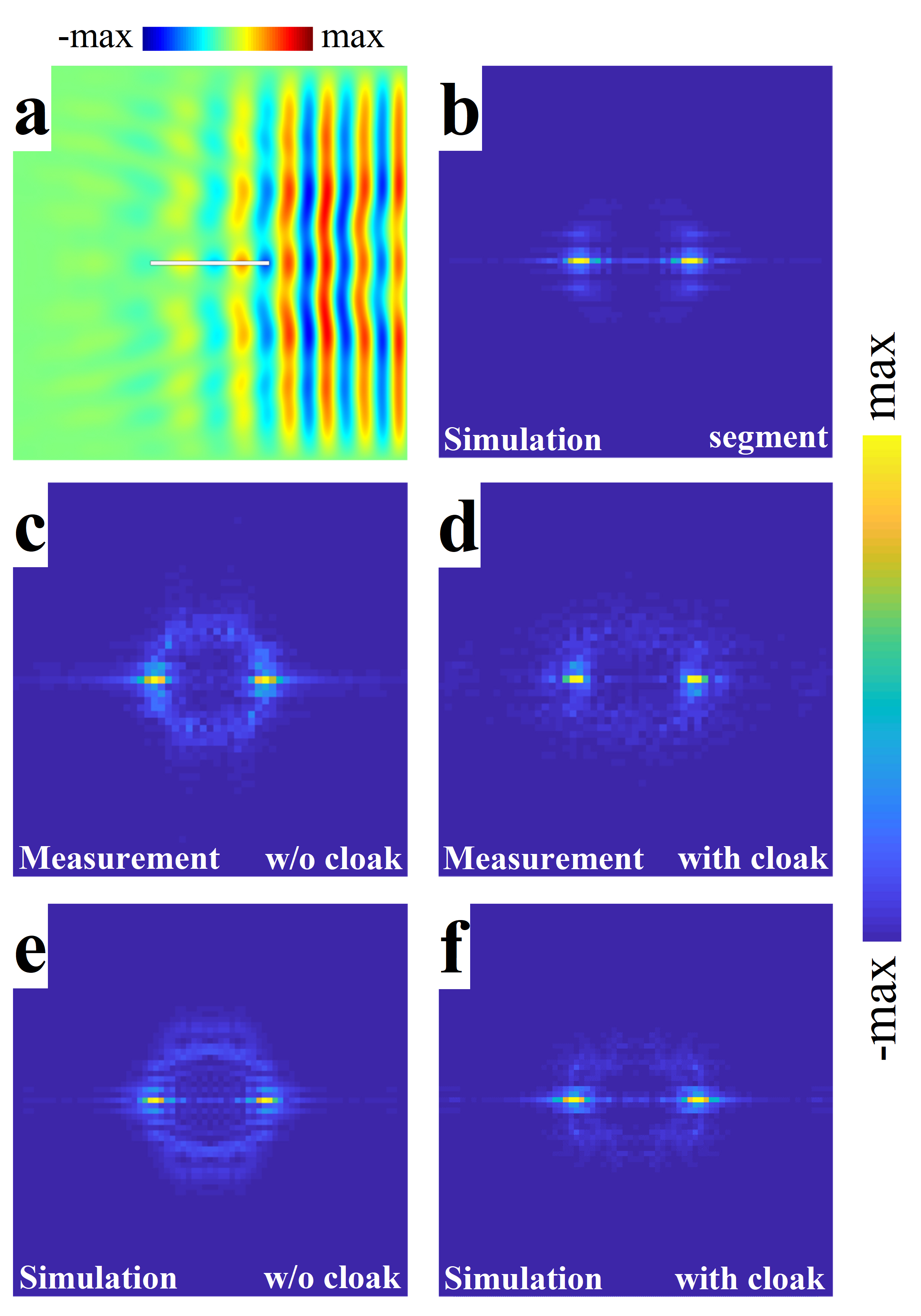}
\caption{(a) Real-space spatial representation of the velocity field amplitude (Gaussian pulse) on a reference plain plate with a transverse segment where stress-free boundaries are applied (no hole, no cloak). (b-f) $k$-space representation of the velocity field amplitude at time step 0.165ms for (b) reference plain plate, (c,e) plate with hole, (d,f) plate with hole and cloak. (a-b, e-f) Full-3D elastic-wave transient simulations; (c,d) experimental results. The strong similarity between (b,d,f) is noted.}
\label{kspace}
\end{figure}

To better quantify the efficiency of the proposed cloak, we investigate the velocity field in the wave-vector ($k$) space. The field amplitude spatial distribution is Fourier transformed and shown in Fig.~\ref{kspace}.
The plain plate is considered for reference (Fig. 5(a)$\&$(b)). The two horizontal narrow strips seen in the k-space (Fig.~\ref{kspace}(b)) show the directivity of the emission and the spectral content of the pulse. The same pattern is recovered for the measurement results of the cloaked hole (Fig.~\ref{kspace}(d)). This is in contrast with the bare hole case, which displays an annular distribution of k-vectors (Fig. 5(c)), showing that the direction of incidence has been lost. This representation allows us to better quantify the ability of the cloak to maintain the initial wavefront.
The simulation results (Fig. 5(e-f)) agree well with the measurement results (Fig. 5(c-d)). The small discrepancy for the experimental invisibility cloak (Fig. 5(d)) comes from the imperfect incident pulse and unwanted residual reflections from the edges perpendicular to the direction of wave propagation.

In conclusion, we have shown that coordinate transformation can be adapted to design a unidirectional cloak for flexural pulses. Our approach only requires modulation of the plate thickness. This considerably simplifies the design of earlier cloaks such as proposed in \cite{PRL_farhat} and experimentally validated in \cite{cloaking_wegner}. Moreover, our approach does not require large contrast in thickness variation \cite{cloaking_darabi}, or beams of varying height atop the plate \cite{colombi2015directional}. Interestingly, the moderate thickness variation in our cloak's design is similar to that of a Maxwell fisheye lens in \cite{lefebvre2015experiments}, that was based upon a conformal route minimizing anisotropy constraints \cite{leonhardt2009broadband}. Finally, thanks to analogies that can be drawn between out-of-plane elastic waves in structured plates and Rayleigh waves propagating in soft soils structured in the first few meters just below the free air-soil surface \cite{brule2014}, we believe our design opens an interesting route to detour surface Rayleigh waves around a prescribed region by artificially structuring a thin layer of soil just below the free air-soil surf..ace, instead of burying concrete columns in soil \cite{colombi2016transformation}.

\begin{acknowledgments}
The authors thank Dr. Leonid Wolfson for preparing the experimental setup. This research was supported by The Israel Science Foundation (Grants No. 1871/15, 2074/15 and 2630/20), the United States-Israel Binational Science Foundation NSF/BSF (Grant No. 2015694 and No. 2021811) and the Youth Foundation Project of Zhejiang Lab (Grant No. 2020MC0AA07). P. S. is thankful to the CNRS support under grant PICS-ALAMO.
\end{acknowledgments}

\bibliography{mybibli}

\end{document}